# Title page:

# From Risk Perception to Behavior Large Language Models-Based Simulation of Pandemic Prevention Behaviors

Youduo Chen*[1], Mingxuan Chen*[1], Lujia Bo[1], Xiaofan Gui[2], Jiang Bian[2], Yi Liu[1,3], Chunyan Wang[#1]

1: School of Environment, Tsinghua University, Beijing 100084, China

2: Microsoft Research, Beijing 100080, China

3: State Key Laboratory of Regional Environment and Sustainability, Tsinghua University, Beijing, 100084, China

* These authors contributed equally to this work.

# Corresponding author: Chunyan Wang, wangchunyan@tsinghua.edu.cn

# Abstract

Individual preventive behaviors constitute a primary line of defense against viral transmission mediated by environmental media (e.g., indoor air and surfaces). Anticipating how individual adjust preventive behaviors is essential for outbreak control, designing environmental containment measures. Yet, behaviors are highly heterogeneous and jointly shaped by disease conditions, environmental and policy context, and individual attributions, leaving key mechanisms insufficiently characterized and limiting the transferability of conventional empirical models across contexts. To address this gap, we propose a large language model (LLM)-based agent framework to simulate individuals' preventive behavior that are environmental media–related. This framework contains (i) a static module for context-conditioned behavior prediction and (ii) a dynamic module that updates perceived risk over time and translates these updates into behaviors trajectories. We evaluate the model using two rounds of survey data from Beijing residents during the COVID-19 outbreak under three data-availability settings: zero-shot, few-shot, and cross-context transfer. Results show that this framework demonstrates robust predictive performance across three settings, effectively capturing the internal logic of preventive behaviors under different data conditions. Behavioral trajectories reveal that perceived risk is a critical mediator between environment and behaviors, and habit formation is the primary determinant for long-term behavioral persistence. Furthermore, this study characterizes a "dual-effect" of community control measures that increasing awareness of the epidemic and increasing the sense of security. This mechanism explains behavioral anomalies such as resistance in over-secured low-risk areas and panic in high-risk settings, which underscore the necessity of adaptive, risk-aligned policy design. This study offers a pragmatic tool to explore resident responses and inform containment strategies for future pandemics before empirical evidence can accumulate.

# 1. Introduction

Evidence from major infectious disease outbreaks, including SARS, H1N1 influenza, and COVID-19, has shown that individual preventive behaviors (IPBs) (e.g., mask wearing, hand hygiene, and environmental disinfection) constitute a first line of defense against viral transmission via environmental media (Mitze et al., 2020; Islam et al., 2020). These behaviors are not merely personal choices but critical interventions that mediate the relationship between environmental exposure and infection risk. IPBs are strongly influenced by government prevention and control policies, as policy strengthen, clarity of guidance, and implementation intensity directly influence public adherence (Van Bavel et al., 2020; Porat et al., 2020). While, IPBs also changes in response to epidemic dynamics, such as shifts in viral transmissibility and transmission risk. Understand how IPBs amid changing policy contexts and epidemic dynamic is essential for designing effective and proportionate policies, reducing inefficient allocation of resources, and improving epidemic control in environmental and public health governance.

However, simulating IPBs is inherently challenging due to the complexity of human decision-making mechanisms. Perceived risk theory posits that behavior is driven by risk perception, a subjective cognitive process rather than objective epidemiological indicators (e.g., strength of policies, virus transmissibility, or lethality) (Brewer et al., 2007; Savadori & Lauriola, 2021; Slovic, 1987). A survey across 10 countries found that risk perception is a significant predictor of prevention behaviors such as mask-wearing and social distancing (Dryhurst et al., 2020). In addition, this mediation effect may vary due to multiple factors, including individual heterogeneity (Daoust, 2020), government prevention and control policies (Hale et al., 2021), and epidemic dynamics (Petherick et al., 2021). Meanwhile, these factors interact with each other, resulting in highly heterogeneous and dynamic prevention behavior patterns. A large number of influencing factors and their intricate interactions pose critical challenges to elucidating the decision-making processes underlying IPBs and to their simulation.

Traditionally, statistical methods (e.g., regression analysis, structural equation models) and machine learning (ML) algorithms are mostly used as behavioral simulation methods. Yet, these methods rely heavily on large-scale, high-quality training data to calibrate parameters and establish reliable predictive relationships, thus hindering their applicability in novel epidemic or shifting policy contexts. For example, a structural equation model developed by Šuriņa et al. (2021) to predict COVID-19 prevention behaviors required over 2,600 survey responses to achieve acceptable model fit, while ML algorithms such as random forests and neural networks also often need thousands of survey responses to avoid overfitting (Caspar et al., 2022). More importantly, these methods show poor cross-context generalization capability when facing novel contexts.

For instance, Roda et al. (2020) show that non-identifiability in model calibration using historical confirmed-case data can lead to very different future projections, undermining generalization to new contexts. This low transferability can be problematic as future epidemics and policy responses are inherently uncertain—relying on models limited to known contexts leaves policymakers without effective tools to anticipate behavioral responses to novel threats.

Large Language Models (LLMs) offer a promising solution for simulating human cognition from natural language, overcoming the data dependency and generalization limitations of traditional models. For instance, Binz and Schulz (2023) demonstrated that LLMs can simulate human cognitive processes in decision-making tasks. Peters and Matz (2024) found that LLMs could infer Big Five personality traits from text with an average correlation of about 0.29 with self-reported scores, comparable to supervised models. Galatzer et al. (2023) showed that LLMs could predict psychiatric functioning from clinical text with strong predictive performance. As IPBs are shaped not only by external conditions, but also by how individuals interpret and psychologically respond to those conditions. Thus, behaviors could be possibly well simulated based on cognitive process.

Furthermore, LLMs can adapt to new tasks under zero-shot or few-shot prompting with minimal task-specific data (Brown et al., 2020; Kojima et al., 2022), and recent work suggests that emotionally loaded prompts can induce modulated anxiety-like responses in LLMs, indicating their potential to capture dynamic affective states relevant to risk-related decisions. Therefore, LLMs are theoretically well positioned to model the subjective, heterogeneous, and context-sensitive determinants of epidemic prevention behaviors that traditional quantitative models often struggle to capture. Yet, these studies offer less clarity about how evolving contextual signals are internally processed into changing behavioral tendencies. In this sense, the key challenge is not whether LLMs can simulate behavior in a dynamic setting, but whether they can represent a theoretically meaningful intermediate that links changing external inputs to behavioral adjustment.

Previous epidemic-behavior models have shown that perceived risk or awareness can serve as an explicit behavioral state linking objective epidemic signals to self-protective actions and disease dynamics (Funk et al., 2009; Poletti et al., 2012). This is precisely the role needed in LLM-based simulation: the model should not only generate plausible behaviors, but also represent how policy changes and epidemic severity alter the perceived risk that mediates behavior. However, the reliability of LLM-based agents remains highly sensitive to prompt design: role instructions, structural constraints, and exemplars can substantially affect consistency, controllability, and the stability of simulated state updates (Shi et al., 2023; Kojima et al., 2022). Existing generative-agent studies have reproduced plausible collective patterns, including disease-related contact dynamics (Williams et al., 2023), but they typically emphasize emergent outcomes rather than a theory-specified cognitive pathway. Therefore, LLMs should be coupled with an explicit behavioral theory, such as perceived-risk theory, rather than used as unconstrained prompts alone.

To address these gaps, this study develops a novel LLM-based agent framework integrated with perceived-risk theory, aiming to model the causal chain from epidemic evolution to updates of risk perception related to viral transmission via environmental media and subsequent behavioral adjustments. We validate the model using empirical COVID-19 survey data from Beijing residents (collected in T1: December 2020 and T2: August 2021 by Li et al., 2025) through three data-availability settings: zero-shot learning (to establish baseline capabilities without reference data), few-shot learning (to enhance accuracy with limited data), and cross-context transfer (to test generalization to novel epidemic contexts using historical data). In addition, we (1) simulate behavioral changes under China's management of COVID-19 as a Class B infectious disease (a dramatic shift from stringent control to comprehensive relaxation); and (2) analyze behavioral changes across 120 scenarios varying in transmissibility, lethality, and community control measures. The remainder of this article is organized as follows: Section 2 describes the model framework, evaluation strategies, scenario settings, and data processing; Section 3 presents the validation results and scenario simulations; Section 4 discusses theoretical and policy implications; and Section 5 concludes the study.

# 2. Methodology

We propose an LLM-based agent framework that incorporates perceived-risk theory to track the causal chain through which epidemic and policy signals update individual risk perception and shape prevention behaviors. In this framework, perceived risk is employed as an explicit intermediate state: the LLM first infers or updates an individual's perceived risk from external epidemic and policy inputs, and this updated perceived risk is then used to generate the likelihood of engaging in specific IPBs. As shown in Figure 1, the framework consists of two modules: (1) a static module that simulates individual attributes and the likelihood of engaging in each IPB under a given epidemic context; and (2) a dynamic module that updates risk perception and IPBs as epidemic conditions and policy settings change over time. We evaluate the framework by comparing its outputs (i.e., likelihood of engaging in IPBs) with survey-based behavioral observations under three data-availability settings: zero-shot, few-shot, and cross-context transfer. After validation, we use the same framework for scenario analyses, including a policy-relaxation scenario and 120 combinations of epidemic and community-control conditions.

In total, this study considered 11 types of IPBs related to viral transmission via environmental media. These IPBs are classified into five categories:

- **Respiratory protection:** mask wearing in community green spaces; mask wearing in elevators.
- **Personal hygiene practices:** hand washing after returning home; closing toilet lids when flushing.
- **Disinfection activities:** regularly disinfecting drain water seals and household surface (routine measures); disinfecting items brought home from outside, including online shopping packages (situational measures).
- **Food safety related measures:** using separate serving utensils at home and canteen; avoiding frozen food purchases.
- **Environmental maintenance practices:** maintaining drain water seals.

For terminological clarity, external environment refers to the combined Pandemic Context and Community Level Control Measures prompt components. Scenario denotes a study-level simulation setting, such as the Policy Relaxation scenario or the multi-dimensional epidemiological scenarios. Condition denotes a specific community-control state within a scenario.

The empirical evaluation considers three community-level control conditions of increasing intervention intensity: Regular Prevention and Control (Regular P&C), Self-Health Monitoring, and Isolation. A fourth condition, No P&C, is introduced only for the subsequent scenario simulations.

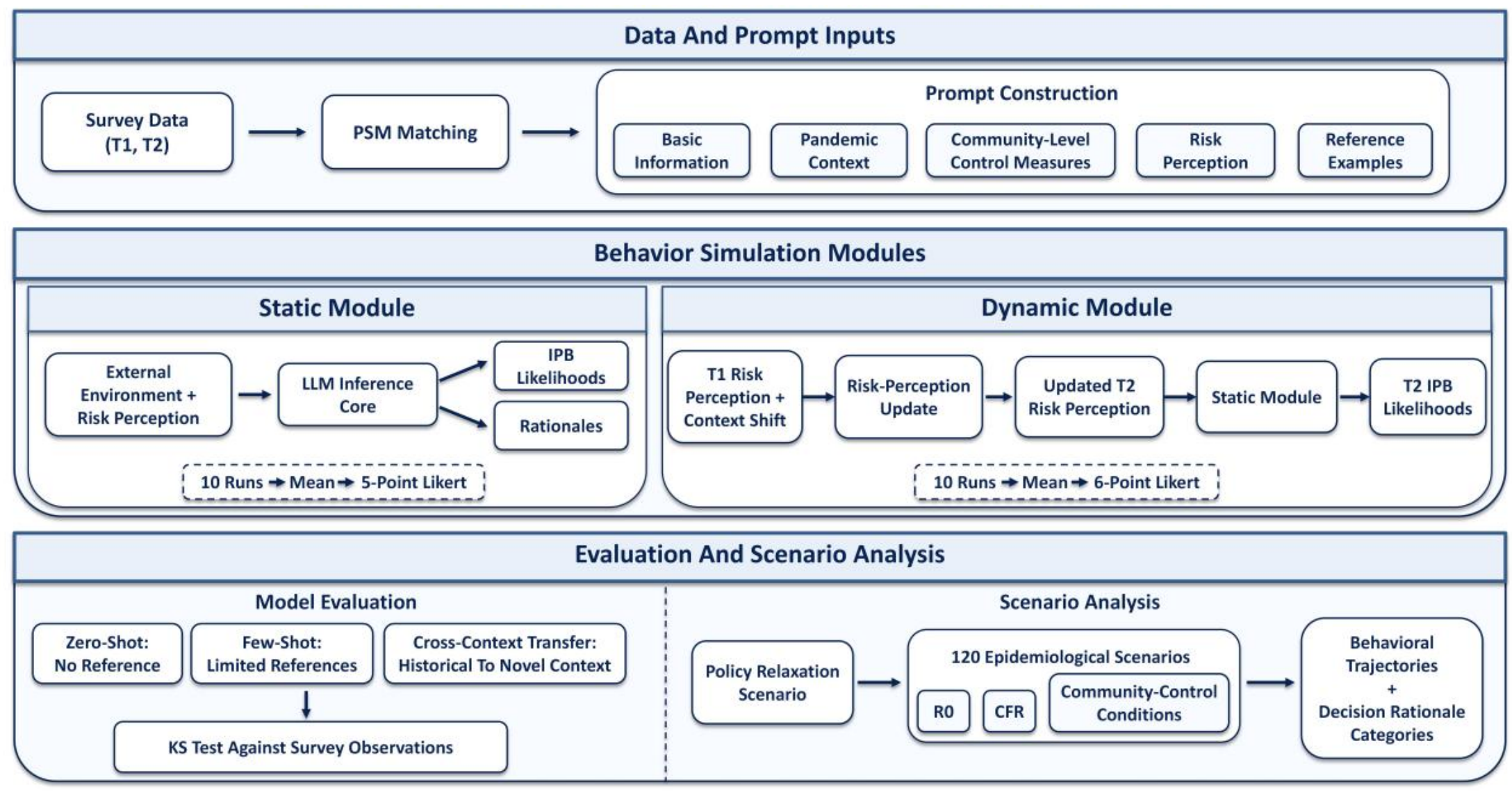

Figure 1. LLM-Based Framework for Preventive Behavior Simulation

## 2.1 Behavior Simulation

### 2.1.1 Static Prevention Behavior Simulation

The technical architecture of the static module consists of five key components:

- **Basic Information:** Establishing individual's identity by incorporating demographic attributes (including age, gender, education level, occupation), enabling the LLMs to conduct individual-specific behavioral reasoning that contains heterogeneous decision-making patterns across different population segments.
- **Pandemic Context:** Providing city-level epidemic information encompassing viral characteristics (i.e., reproduction number ($R_0$) representing transmissibility, case fatality rate (CFR) representing lethality, and transmission pathways) and epidemic severity information (i.e., confirmed cases, fatalities, and overall outbreak trends).
- **Community Level Control Measures:** Representing community-level interventions that may influence residents'preventive decisions, including testing and screening requirements, access management, mobility restrictions, health monitoring, and isolation arrangements. The complete inventory of measures, coding rules, and corresponding prompt descriptions for each empirical and simulated condition are provided in the Supplementary Information.
- **Environmental Risk Perception:** Establishing the psychological cognitive state and guiding the LLMs to incorporate subjective risk cognition into behavioral decision-making processes.

- **Task Setting:** Instructing the LLMs to output: (i) the execution likelihood (continuous value between 0 and 1) for the considered 11 IPBs and (ii) rationales underlying these likelihood results.

The Pandemic Context and Community Level Control Measures constitute the external environment influencing residents' behavioral decisions. The environmental risk perception is obtained from the empirical surveys adopted from Li., et al (2025), with values measured by 6-point Likert scales (1=unclear/unconcerned to 6=extremely concerned). To mitigate the inherent stochasticity of LLMs (Atil et al., 2024), we execute ten simulation rounds for each behavior and calculate the mean values as the final behavior's likelihoods. Moreover, to align our predictions with the empirical survey data (Li., et al, 2025), we discretize the simulated behavior likelihoods into a 5-point Likert scale (1=Never adopted to 5=Always adopted) using equidistant thresholds.

**2.1.2 Dynamic Prevention Behavior Simulation**

The dynamic module simulates changes in individual risk perception and associated IPBs throughout external environment evolution from period T1 (December 2020) to T2 (August 2021). The risk perception updates are achieved via a prompt engineering architecture analogous to that of the static simulation module:

- **Basic Information:** The same individual identity information as used in the static simulation module to anchor individualized reasoning.
- **Pandemic Context Shift:** Representing the change in epidemic conditions from T1 to T2, including shifts in viral characteristics (e.g., transmissibility, lethality) as well as changes in epidemic severity (e.g., case levels, fatalities, and overall outbreak trends).
- **Changes in Community-Level Control Measures:** Representing transitions in the status or intensity of the community-level measures defined in the static module, including their strengthening, maintenance, relaxation, or withdrawal between T1 and T2. The detailed transition settings and prompt templates are provided in the Supplementary Information.
- **Task Setting:** Instructing the LLM to output: (i) the updated environmental risk perception level for T2 based on the individual's perception level at T1 and the other three input components; and (ii) the rationale underlying this update.

Same as the static module, we also calculate the mean values of ten iterations as final risk perception levels and discretize them into 6-point Likert scales (1=unclear/unconcerned to 6=extremely concerned) using equidistant thresholds. The updated risk perception levels are then used as inputs of the static module to simulate the IPBs for period T2.

## 2.2 Model Evaluation

The framework's performance under varying data-availability conditions is evaluated against the empirical survey data from Li et al. (2025), as summarized in Table 1.

### 2.2.1 Model Evaluation Strategies

**Strategy 1 Zero-shot evaluation.** As a fundamental paradigm in language model evaluation (Brown et al., 2020; Kojima et al., 2022), zero-shot evaluation tests whether LLMs can leverage their pre-trained knowledge to reason about human behavioral responses solely on scenario descriptions. We use this strategy to assess whether the proposed framework could simulate IPBs without relying on any survey-based behavior information as references.

**Strategy 2 Few-shot evaluation.** Few-shot learning enhances LLMs' task understanding by providing a small number of input-output pairs as references (Liu et al., 2023; Min et al., 2022). For static simulation, we randomly select one-third of the T1 samples as references and embed their IPBs information in prompts to predict IPBs of the remaining samples. For dynamic simulation, we use the Regular Prevention and Control (P&C) condition in T2 as a case study and apply the same data selection approach as in the static simulation. This strategy enables us to assess the framework's ability to improve simulation accuracy through contextual learning with limited examples.

**Strategy 3 Cross-context transfer.** This strategy assesses the framework's capacity to generalize from historical data to predict behaviors in novel epidemic contexts. For static simulation, we use T1 data as reference to predict behaviors in T2. For dynamic simulation, we use the behavioral data under the Regular P&C condition for the T1-to-T2 transition and evaluate the model outputs on the other two conditions (i.e., Isolation and Self-Health Monitoring conditions). This strategy assesses whether behavioral patterns learned in one context can support predictions in different contexts and conditions.

Table 1 Model Evaluation Strategies

| | **Static Behavior Simulation** | **Dynamic Behavior Simulation** |
|---|---|---|
| **Zero-shot** | Reference: None<br>Test: T1 | Reference: None<br>Test: T1 to T2 (i, s, r) |
| **Few-shot** | Reference: 1/3 T1<br>Test: 2/3 T1 | Reference: 1/3 T1(r) to T2(r)<br>Test: 2/3 T1(r) to T2(r) |
| **Cross-context Transfer** | Reference: T1 (i, s, r)<br>Test: T2 (i, s, r) | Reference: T1 to T2 (r)<br>Test: T1 to T2 (i, s) |

Notes: i, s, and r refer to Isolation, Self-Health Monitoring, and Regular Prevention and Control (P&C) conditions, respectively. "T1 to T2" denotes a simulated transition from the T1 context (December 2020) to the T2 context (August 2021), rather than a subtraction operation.

It should be noted that the behavior simulation is advanced to cross-context transfer strategy only after passing few-shot strategy, which establishes the proposed framework's basic learnability from reference examples. Likewise, successful static simulation is required before dynamic simulation.

#### 2.2.2 Model Performance Evaluation Methods

We employ Kolmogorov-Smirnov tests (KS tests) to compare the distributions of the simulated and observed behaviors. The KS statistic serves as a measure of the distance between the two cumulative distribution functions, where a lower value indicates a higher degree of similarity and thus better model accuracy (Massey, 1951). We establish a significance threshold of 0.001. A p-value exceeding this threshold ($p > 0.001$) indicates a failure to reject the null hypothesis, suggesting that the simulated behavior distributions do not significantly differ from the observed distributions.

### 2.3 Scenario Settings

#### 2.3.1 Settings for Policy Relaxation Scenario

In December 2022, China adjusted its COVID-19 control regime through the official notice Notice on Further Optimizing and Implementing COVID-19 Prevention and Control Measures, commonly referred to as the "New Ten Measures" (https://www.nhc.gov.cn/xcs/zhengcwj/202212/15457b5897ba42c991e3ee3b1f1f4bed.shtml). On December 25, 2022, the National Health Commission further announced that it would no longer publish daily epidemic information. These policies introduced significant changes relative to earlier strict control requirements. We operationalize this policy shift by modifying the prompt contents for the Pandemic Context and Community Level Control Measures components, and define this setting as the Policy Relaxation scenario to simulate changes in IPBs in response to shifts in the external environment. This scenario is selected as it represents a natural temporal extension of our empirical survey data, testing the framework's forward-looking predictive capabilities under a drastically altered policy landscape.

For the Pandemic Context, the viral transmissibility is set as $R_0 = 10.0$ (vs. ancestral strain $R_0 \approx 2.2$ (Li et al., 2020)) and lethality is set as CFR = 0.05% (Chenchula et al., 2023; Xia et al., 2024) to reflect the Beijing epidemic wave following the December 2022 policy shift, during which BA.5.2 and BF.7 were the dominant circulating Omicron subvariants. Because official daily case and death reporting was discontinued after December 25, 2022, the epidemic severity information within this Pandemic Context component was described using qualitative descriptors (e.g., “rapid widespread infection” and “surging case numbers”) rather than numerical counts.

For the Community Level Control Measures component, the Policy Relaxation scenario was operationalized by changing the status of all measures defined in the framework from active to cancelled or inactive. This setting represents the broad withdrawal of community-level interventions following the December 2022 policy shift. The measure-specific coding and complete prompt configuration are reported in the Supplementary Information.

Finally, to ensure that the LLMs' reasoning remained anchored in the contemporaneous governance environment, we embedded a concise summary of the post-December 7 policies within the Pandemic Context narrative.

#### 2.3.2 Settings for Various Epidemiological Scenarios

The Policy Relaxation scenario only captures changes in IPBs in response to a specific external environment setting. To more fully characterize how the external environment shapes individuals’ behavioral decision-making, we construct a multi-dimensional scenario framework by systematically varying three dimensions that jointly define the epidemic context: two viral characteristics ($R_0$ and CFR) and the community level control conditions.

Specifically, $R_0$ spans six values (0.8, 2, 3, 5, 7, 10); CFR is configured at five levels (0.1%, 0.5%, 1.5%, 3.0%, 5.0%). The combinations of $R_0$ and CFR cover a broad spectrum of respiratory viral outbreaks, encompassing epidemiological characteristics of common respiratory pathogens. The community-level control condition comprises four levels of increasing intervention intensity: No P&C, Regular P&C, Self-Health Monitoring, and Isolation.

In total, 120 epidemic scenarios are generated (6 × 5 × 4). Each scenario is operationalized by modifying the Pandemic context and Community Control Measures components in the prompts, enabling LLMs to conduct behavioral reasoning under the specified settings (examples are given in SI).

## 2.4 Decision rationale analysis

In addition to numerical outputs, the LLMs are required to provide concise decision rationales for each simulated behavioral likelihood and each updated risk-perception score in the dynamic module. To interpret heterogeneous behavioral trajectories, rationales from individuals with relatively high or low simulated behavioral likelihoods for focal behaviors are extracted and coded into theoretically informed categories, including perceived risk, habit formation, behavioral cost, convenience, official guidance, and perceived protection from community-level control measures. When a rationale contains multiple explanations, all relevant categories are recorded. The frequency of each category is then summarized to identify dominant explanation patterns underlying heterogeneous behavioral trajectories.

## 2.5 Data Collection and Processing

The empirical data used in this study are obtained from two questionnaire surveys conducted in Beijing during the COVID-19 pandemic by Li., et al (2025): Round 1 ( December 2020 T1; sample size n = 980) and Round 2 (R2, August 2021 T2; sample size n = 120). Both survey rounds covered the three empirical community-level control conditions defined in Section 2.1. Four types of information of each respondent are used in this study: (1) demographic characteristics, including age, gender, education level, and occupation; (2) self-reported prevention behaviors, measured for 11 IPBs using a 5-point Likert scale; (3) risk perception, measured on a 6-point Likert scale across four viral transmission pathways via environmental media and ten specific exposure scenarios; and (4) perceived intensity of community-level control measures experienced by the respondent. These four types of information are used to: (1) provide observed behavioral outcomes when evaluating the proposed framework; and (2) supply key individual- and context-level inputs for prompt construction. Detailed questionnaire items and variable coding are provided in the Supplementary Information.

As the two survey rounds sampled different individuals rather than tracking the same people over time, direct comparison between T1 and T2 may be confounded by differences in sample composition. To improve comparability, we apply propensity score matching (PSM), a standard technique in observational behavioral and public-health research (Rosenbaum & Rubin, 1983; Austin, 2011; Chaudhuri et al., 2022; de Lusignan et al., 2021). Specifically, one-to-one nearest-neighbor matching is performed using demographic variables, including age, gender, education level, and occupation. This procedure yields 120 matched pairs, which served as the unified analytic sample for model evaluation. The static module used the matched observations at each time point for cross-sectional behavior prediction, whereas the dynamic module used the paired T1-T2 structure to evaluate changes in risk perception and preventive behaviors over time.

## 2.6 Prompt Construction and Simulation Implementation

Structured prompts are constructed from the components described in Section 2.1. This section summarizes how survey variables and scenario parameters are converted into prompt-ready inputs, how reference examples are embedded under different evaluation strategies, and how model outputs are processed.

**Personal profile prompt construction**

We adopt a first-person prompt design to represent each respondent as a simulated resident, because prior studies suggest that persona-based prompting can improve the consistency of individualized behavioral reasoning (Argyle et al., 2023; Hu & Collier, 2024). Demographic characteristics collected from the surveys are mapped directly to the Basic Information component of the prompt. To enhance the realism of the first-person perspective, anonymous respondent IDs are replaced with virtual Chinese names generated from a standardized name corpus. In addition, categorical age intervals in the survey are converted into specific ages by random sampling within the reported range (e.g., “20–30” was converted to a specific age such as 24), so that the prompt contained concrete rather than interval-based persona information.

**External-environment prompt construction**

The Pandemic Context component is written from scenario-specific viral characteristics ($R_0$, CFR, and transmission pathways) and epidemic severity information. For empirical evaluations, the Community Level Control Measures component is constructed from the survey-based perceived intensity of community-level measures. Because the survey did not record the full administrative policy file for each community, we averaged these perceived-intensity responses within each community to generate the community-level input. For scenario simulations, this component is instead constructed from the researcher-specified control condition (No P&C, Regular P&C, Self-Health Monitoring, or Isolation).

**Risk-perception prompt construction**

The questionnaire measured respondents' perceived infection risk related to viral transmission via environmental media across four transmission pathways and ten exposure scenarios. For the static module, these responses are discretized into a 6-point Likert scale and used to construct the Risk Perception component. For the dynamic module, the T1 risk-perception score is used as the initial state, and the prompt asks the LLM to update this state under the T1-to-T2 change in the external environment.

**Reference-example construction**

The 11 self-reported prevention behaviors measured in the survey are not used as direct prompt inputs in the zero-shot setting; instead, they serve as empirical observations against which simulated outputs are evaluated. In the few-shot and cross-context transfer settings, selected behavioral records are embedded as reference examples in the prompt following the evaluation design described in Section 2.2.

**Output processing**

After each prompt response is returned, numerical behavior likelihoods and risk-perception scores are parsed and converted to the corresponding Likert scales as described in Section 2.1. The accompanying rationales are used only for the decision-rationale analysis described in Section 2.4 and are not treated as prompt inputs.

## 2.7 Model Implementation

The main-text simulations were implemented using Python 3.8 with the OpenAI API for accessing GPT-4o (model version: gpt-4o-2024-08-06). To assess the robustness of the framework to model choice, we additionally repeated the simulation workflow using GPT-5, with the corresponding results reported in the Supplementary Information. API requests were executed asynchronously with exponential backoff for rate-limit handling and processed in batches of 10 concurrent requests.

# 3. Results

## 3.1 Model Evaluation and Comparison

We compared the absolute levels and temporal patterns of behavioral likelihood (static module and dynamic module) and risk perception (dynamic module) with those reported by Li et al. (2025) for model evaluation.

### 3.1.1 The Static Module Performance

Results of the static module under three data availability strategies are presented in Figure 2. Overall, the majority of IPBs failed to reject the KS tests ($p>0.001$), indicating strong performance of the static model in simulating behaviors. Specifically, **Strategy 1 (Zero-Shot Learning)** accurately simulated eight out of eleven behaviors under T1 with Regular P&C, with all p-values exceeding the 0.001 threshold (c.f., Figure 2A). **Strategy 2 (Few-Shot Learning)** substantially improved model performance, with nine of the eleven behaviors successfully modeled (c.f., Figure 2B). "Disinfecting online purchase packaging" and "maintaining drain water seals" were well simulated in Strategy 2 but poorly simulated under Strategy 1, whereas "regularly disinfecting home environment" showed the opposite pattern, succeeding under Strategy 1 but failing under Strategy 2. **Strategy 3 (Cross-Scenario Transfer)** demonstrated the LLMs' capacity to transfer patterns learned from historical data to novel epidemic scenarios, with seven behaviors meeting the accuracy threshold ($p>0.001$) under T2 across all three control measures (c.f., Figure 2C-E). Notably, both mask-wearing behaviors (in community green spaces and in elevators) showed poor performance in this cross-scenario evaluation ($p<0.001$), likely suggesting that these behaviors are highly dependent on specific external environmental contexts and the implementations of community-level control measures.

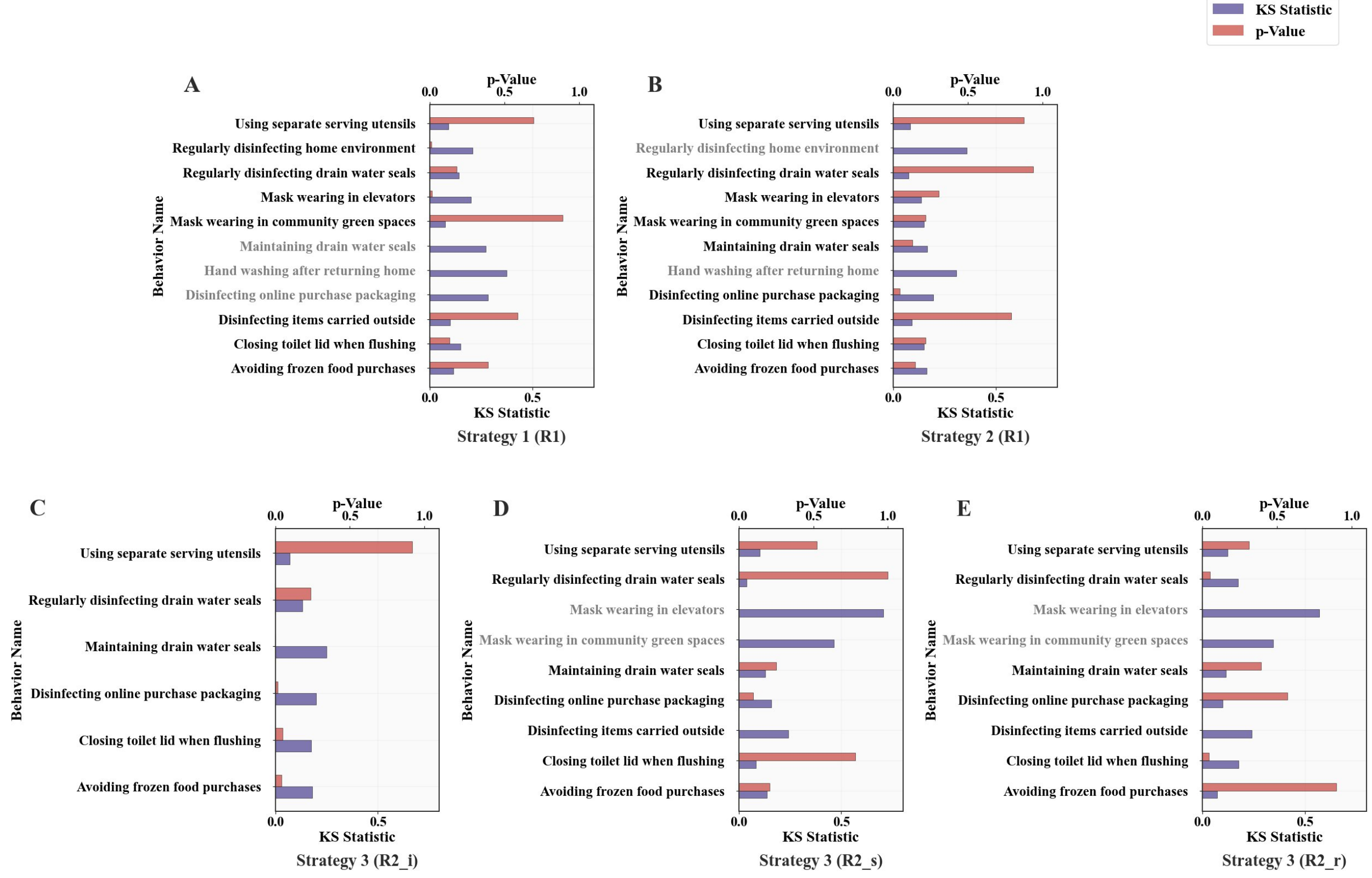

Figure 2. Evaluation of Static Prevention Behavior Simulation.

Notes: Black text indicates behaviors failed to reject the KS tests (p-values > 0.001), while gray text indicates behaviors passed the KS tests. Figure 2A shows Strategy 1 evaluation results for T1 Regular P&C. Figure 2B presents Strategy 2 evaluation results for T1 Regular P&C. Figure 2C, 2D, and 2E display Strategy 3 evaluation results for T2 across three control conditions (*i* = Isolation, *s* = Self-Health Monitoring, *r* = Regular P&C), with T1 used as training data.

### 3.1.2 The Dynamic Module Performance

The dynamic module was also evaluated across the three strategies for behaviors that passed the static module (Figure 3). **Strategy 1 (Zero-shot)** showed varying model performance under T2 across the three community-control conditions: all five behaviors under the Isolation condition were accurately simulated, whereas five behaviors under the Self-Health Monitoring condition and three behaviors under the Regular P&C condition were accurately simulated, respectively (Figure 3A-C). Notably, mask-wearing behaviors (in community green spaces and in elevators) also showed poor performance, mirroring the findings from the Static Module performance under Strategy 3. **Strategy 2 (Few-Shot)** achieved significant performance improvement from T1 to T2 under Regular P&C, with all seven behaviors passing the evaluation ($p>0.001$) (Figure 3D). **Strategy 3 (Cross-Scenario)** demonstrated robust cross-scenario transfer capability: five of six behaviors met the accuracy threshold ($p>0.001$) under the Isolation condition (Figure 3E), while six of seven behaviors were well simulated under the Self-Health Monitoring condition (Figure 3F). Under both condition, the "closing toilet lids when flushing" behavior, which performed well in the static simulation (Figure 2C-D), showed poor performance ($p < 0.001$) in the dynamic simulation (Figure 3E-F).

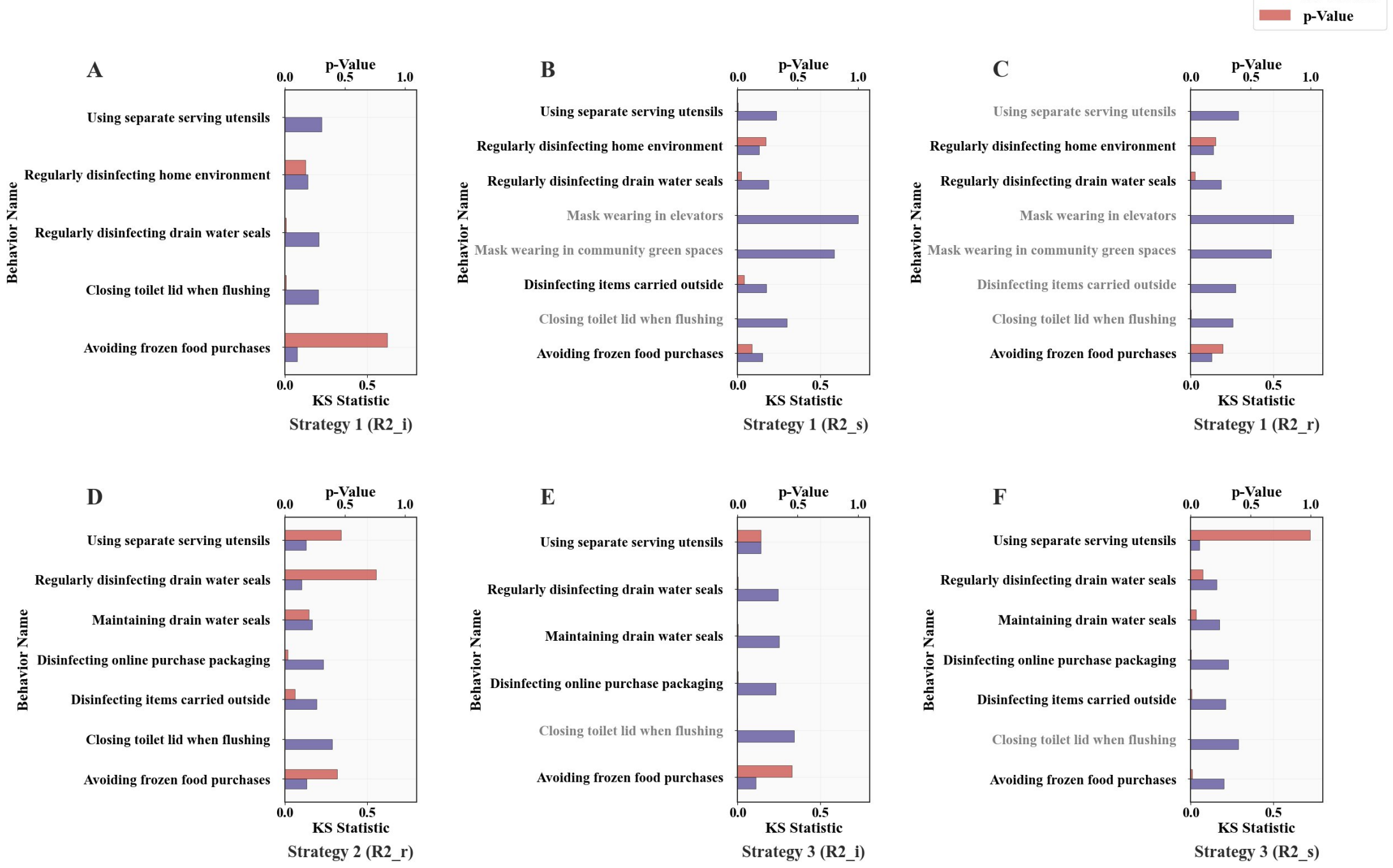

Figure 3. Evaluation of Dynamic Prevention Behavior Simulation.

Notes: Refer to Figure 2 for text color and statistical definitions. Panels A–C: Strategy 1 (*i*, *s*, *r*); Panel D: Strategy 2 (T2 Regular P&C); Panels E–F: Strategy 3 (*i*, *s*) using Regular P&C as

reference.

## 3.2 IPBs Changes under China's COVID-19 Policy Relaxation

Pronounced heterogeneity in residents' risk perceptions and IPBs was identified during China's COVID-19 policy shift in December 2022 (Figure 4). Compared with the sharper decrease from R1 (4.16 in December 2020) to R2 (3.73 in August 2021), the overall risk perception moderated to 3.62 in this scenario, reflecting a sense of "pandemic fatigue" that led to a decline across most IPBs. The avoidance of frozen food purchases exhibited the most substantial and continuous decline, with its behavioral likelihood falling from 3.61 in R1 to 3.20 in R2, and further to 2.99 in this scenario—a total reduction of 17.2%. In contrast, regular disinfection of drain water seals demonstrated a distinctive counter-trend, increasing from 3.31 in R1 to 3.54 in this scenario. Notably, this was the only IPB that maintained an upward trajectory following the policy relaxation (Figure 4C).

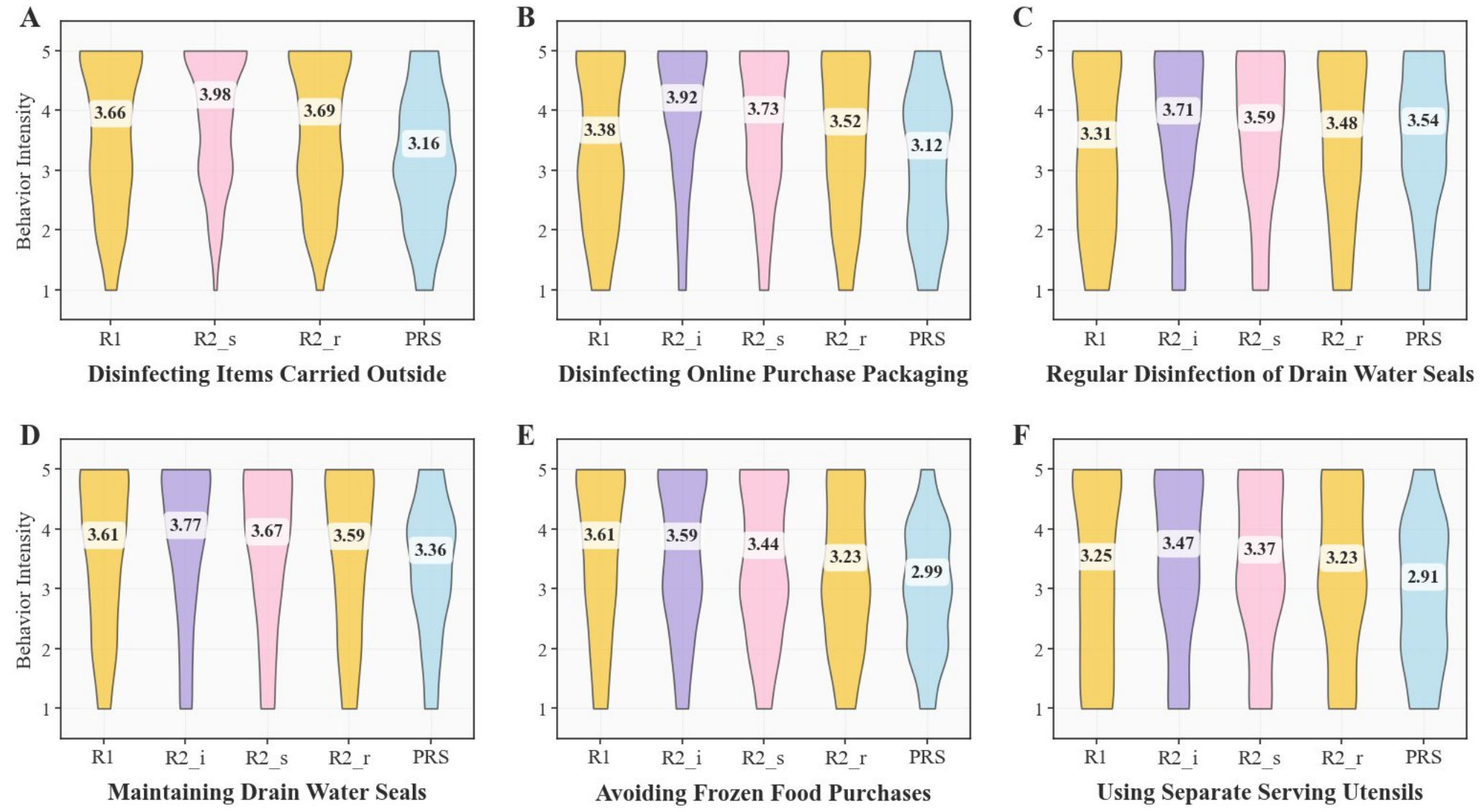

Figure 4. Evolution of Prevention Behavior Likelihood Distribution

Notes: R1 and R2 represent empirical data distributions, where R1 refers to the Regular P&C. R2 encompasses three conditions (i = Isolation, s = Self-Health Monitoring, r = Regular P&C). PRS (Policy Relaxation Scenario) represents simulation results under the No P&C condition. All annotated values indicate the means of behavioral likelihoods.

Analysis of the decision rationales suggests that these heterogenous trends were driven by the interplay between risk perception and habit formation. While previous studies show that habit formation timelines vary significantly (from 18 to 254 days), the intervals in our study—8 months between R1 and R2, and 16 months between R2 and this scenario—provided sufficient duration for habit formation. Specifically, among residents with high behavioral likelihoods for frozen food avoidance, 68% of them cited a diminished risk perception, heavily influenced by "updates of official scientific guidance" regarding frozen product safety. This is consistent with the simulated specific risk perception, which fell from the empirical 4.43 in R1 to a simulated 3.42 in this scenario. Furthermore, 13% noted that the economic cost of such avoidance contradicted ingrained consumption habits. Conversely, among residents with high behavioral likelihoods for drain disinfection, 40% of them referenced a resurgent risk perception concerning viral propagation through drainage systems. This corresponds with the simulated rebound in risk perception concerning buildings's anitary plumbing systems and wastewater to 3.66, following empirical values of 3.79 in R1 and 3.60 in R2. Combined with the 34% who emphasized "low cost" and "ease of maintenance," these factors facilitated the transition from temporary compliance to a sustained habit.

## IPBs changes under Various Epidemiological Scenarios

This study focuses on regular drain disinfection behaviors in the multi-dimensional scenario simulation, as it is the only IPB that shows a distinct upward trend. This behavior also presents a trade-off between epidemic prevention and environmental protection, as the disinfectants used are directly discharged into domestic wastewater systems.

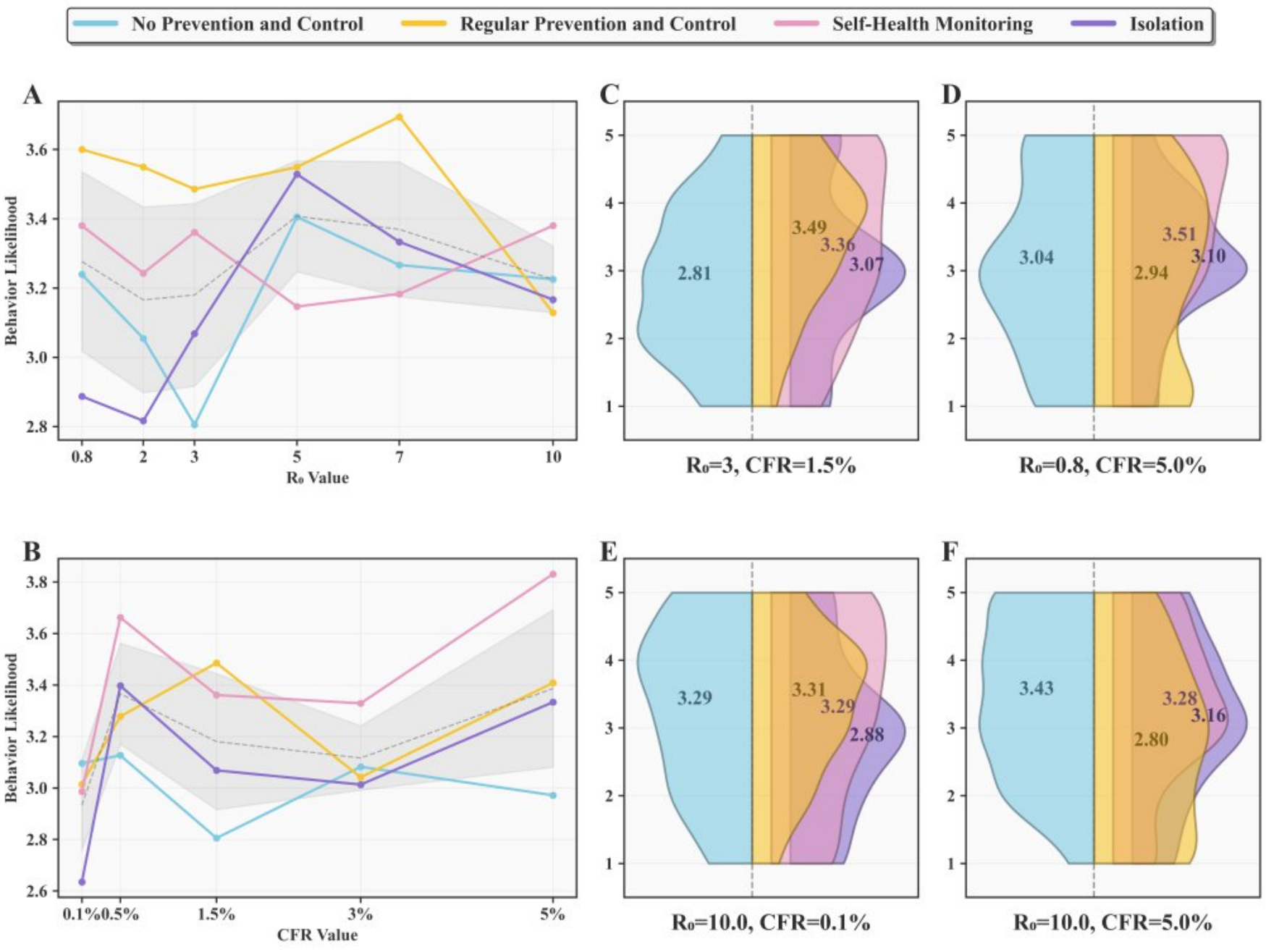

Figure 5. Evolution of Regularly Disinfecting Drain Water Seals Under Multi-Dimensional Epidemic Scenarios

Notes: Panel A illustrates how mean behavior likelihood varies with $R_0$ across four control scenarios when CFR=1.5%. Panel B demonstrates how mean behavior likelihood changes with CFR across four control measures when $R_0$=3. In panels A and B, gray dashed lines indicate the overall average likelihood across all four control measures, while the shaded areas represent the variability range (±1 standard deviation) of behavior likelihoods among the four control measures.Panels C-F display simulated behavior likelihood distributions under different scenarios, with annotated numbers indicating mean behavior likelihoods for each scenario.

As shown in Figures 5A and B, behavior likelihood generally exhibits a fluctuating upward trend as transmissibility and mortality rates increase which particularly evident under the Isolation condition. For instance, when the CFR rose from 0.8% to 10.0%, behavior likelihood under Isolation increased from 2.89 to 3.17. Similarly, when R0 increased from 0.1 to 5.0, the likelihood rose from 2.64 to 3.33. This observation reinforces the perceived risk theory, confirming the causal chain where epidemic risk signals update individual risk perception, which in turn drives behavior likelihood.

However, the impact of control measures on behavior likelihood is more complex and may produce unintended consequences. As illustrated in Figures 5C and E, control measures show a clear promotional effect on behavior, with Regular P&C increasing likelihood by 24.2% and 0.6% relative to No P&C. Yet, as control strengthen further increases, behavior likelihood declines rapidly, with Isolation decreasing by 12.0% and 13% compared to Regular P&C. Notably, under extreme high-risk conditions (Figure 5F, R0=10.0, CFR=5.0%), behavior likelihood under No P&C reached 3.43, substantially higher than under the other three control conditions. Within the perceived risk framework, this may reflects the dual utility of control measures as they simultaneously heighten risk perception through warning signals and enhance residents' sense of security, which ultimately reduces the perceived necessity for IPBs.

Notably, in conditions with both minimal risk (Figure 5B, CFR=0.1%) and extreme risk (Figure 5F), the behavior likelihood under No P&C consistently exceeded that of the various managed conditions. This observation may suggest a dual-faceted impact of residents' sense of security enhanced by community control. In low-risk conditions, the presence of control measures fosters a sense of over-security. Conversely, under high-risk conditions, the absence of such measures creates a deficiency of security driving individual behavior to peak intensity.

# 4. Discussion

## 4.1 Data Efficiency and Practical Utility of LLMs in Behavioral Simulation

This study demonstrates that LLMs can simulate IPBs in the absence of prior behavioral data as reference, and achieve enhanced precision with minimal data input. For instance, in the static prevention behavior simulation, when no reference data is provided (Strategy1, Zero-shot), LLMs can predict IPBs based on their pre-trained knowledge, reaching a predictive accuracy of 72.7%. Meanwhile, the introduction of limited reference data (Strategy 2, Few-shot) substantially enhances model performance, increasing accuracy to 81.8%. In contrast, traditional statistical methods usually require substantial training data to establish parameter values and validate model structures. Similar findings were supported by existing studies, for example, Xu & Wang (2025) found that GPT-4o achieved comparable simulation performance for residents' travel behaviors using 98% less data than gradient boosting models.

Additionally, our model's utility is demonstrated by its success in predicting behavioral patterns in novel epidemic contexts using historical data. This utility evident in Strategy 3, where the model successfully simulated behaviors across unknown epidemic and policy conditions based on historical few-shot. The only exceptions were mask-wearing in the static simulation and "closing toilet lids when flushing" in the dynamic simulation. Such performance suggests that LLMs can capture the underlying behavioral logic, rather than merely replicating surface-level data patterns from few-shot. From a practical standpoint, this capability enables researchers and policymakers to anticipate public reactions to emerging health crises and design suitable policy, even when real-time empirical data are scarce.

In the present study, perceived-risk theory serves this role by providing a structured and interpretable mechanism to track how epidemic and policy signals update individual risk perception, and how these changes in perceived risk, in turn, shape the adoption or abandonment of prevention behaviors.

## 4.2 Interpretation of Behavioral Changes

By coupling the LLM with an explicit perceived-risk theory framework, this study provides a interpretable mechanism to track how epidemic and policy signals update individual behaviors. While risk perception was the primary programmed driver, the simulation results further reveal that habit formation acts as the decisive factor for long-term persistence.

The unique growth in drain disinfection illustrates this interplay. Although general pandemic concern faded, the simulated risk perception regarding buildings' sanitary plumbing systems and wastewater remained resilient, providing the necessary motivational stimulus. More importantly, the results suggest that this low-cost behavior successfully transitioned into a habit due to its low entry barrier and ease of integration into daily routines. In contrast, frozen food avoidance failed to habituate; once the perceived risk declined, the high economic and lifestyle costs became insurmountable barriers, leading to the behavior's abandonment.

These findings validate the necessity of constraining LLMs with explicit behavioral theory. The framework not only tracks the direct impact of risk perception on behavior but also allows for the inference of deeper mechanisms, such as habit formation, that determine behavioral trajectories. This interpretable mechanism moves beyond unconstrained generation to offer a robust, theory-grounded explanation.

## 4.3 Environmental impacts of IPBs

An often-overlooked dimension of pandemic response is the environmental consequences of IPBs. Taking the regular drain disinfection as an example, when its behavior likelihood increased from the 2.62 (R0=3.0, CFR=0.1%, under Isolation condition) to 3.83 (R0=3.0, CFR=5.0%, under self-health monitoring condition), the annual consumption of disinfectants (at 500 mg/L available chlorine) per capita is estimated to increase by 11.2 liters, leading to an increase of annual disinfection by-product (DBP) generation by approximately 13.4 mg per capita. For Beijing with 22 million residents, this behavioral change could result to an additional annual discharge of over 244,518 tons of disinfectants and 292 kilograms of DBPs into wastewater systems. These DBPs would exert a significant environmental burden, posing a severe threat to aquatic ecosystems (Cui et al., 2021). These environmental consequences should be taken into account when guiding residents toward "right-sized" prevention. Encouraging vigilance during high-risk phases while curbing over-prevention during low-risk phases may help maintain public safety without imposing unnecessary chemical burdens on wastewater systems and aquatic ecosystem.

## 4.4 Policy Implications

BDistinct from previous research using LLMs to simulate individual behavior, our study validates model effectiveness through empirical data, providing robust evidence for the practical applicability of LLM-based behavioral simulation in the field of environmental sociology.

The multi-dimensional epidemiological conditions simulation suggests that community level control measures should be calibrated precisely and flexibly according to viral characteristics and another epidemic context. When such measures are appropriately aligned with the severity of epidemic context, they may both enhance residents' awareness of the epidemic and strengthen their sense of security (Wong & Jensen, 2020). However, stricter control is not necessarily associated with better behavioral outcomes. For instance, the scenario results shown in Figure 5C indicate that behavior likelihood declines as control strength increases, although it remains higher than in the No P&C scenario. This finding suggests that stricter control does not necessarily lead to stronger individual preventive behavior. Therefore, rigid or “one-size-fits-all” strategies should be avoided, as they may reduce intervention effectiveness or generate unintended consequences.

More specifically, the negative consequences of a mismatch between risk level and control level can be observed under both low- and high-risk conditions. When risk levels are low but community control measures are overly stringent, residents may experience psychological reactance and reduced compliance, leading to a marked decrease in behavior likelihood (Díaz & Cova, 2022). This pattern is reflected in Figure 5B, where behavior likelihood under the No P&C condition was significantly higher than under the other control conditions during periods of extremely low transmissibility or mortality. Conversely, under high-risk conditions, the lack of community control may heighten residents' risk perception and even trigger panic response, thereby driving more intensive IPBs (Wardman, 2020). As shown in Figure 5F, under high transmissibility and mortality conditions, behavior likelihood under No P&C condition was substantially higher than that observed under the other three control conditions. These findings further underscore the importance of aligning community-level interventions with actual risk conditions.

# 5. Conclusions

This study establishes a novel computational framework that validates the advantages of LLMs coupled with theory in simulating IPBs under data-scarcity. The model’s ability to generalize across scenarios and provide interpretable reasoning was validated through case studies, offering a valuable decision-support tool for early-stage pandemic policy development. The main conclusions are as follows:

(1) LLMs possess strong zero-shot and few-shot predictive capabilities. This research confirms that even minimal sample integration significantly enhances performance and cross-scenario generalization, underscoring the model's ability to learn the underlying logic of human behavior from few shot.

(2) Simulations further prove that perceived risk serves as a mediation effect between environment and behavior. Meanwhile, it was found that habit formation, as a decisive factor for long-term behavioral persistence, can be uncovered by LLMs. These two factors together drive the behavioral simulation, providing a robust, theory-based explanation for the results.

(3) Integrating perceived risk theory with multi-scenario simulations, this study proposes a dual effect of community control on risk perception: increasing awareness of the epidemic and increasing the sense of security. The findings clarify two behavioral anomalies, including resistance caused by over-security in low-risk environments and individual panic due to policy absence in high-risk contexts, highlighting the need for policies that are consistent with epidemic context and adaptive in nature.

Yet, several limitations remains. The validation is limited by a small sample of Beijing residents (n=120), which may affect prediction accuracy in specific community contexts. Additionally, our pilot-scale knowledge base provided only modest gains; future work should focus on building comprehensive behavioral databases to better integrate expertise with LLMs. Finally, while our prompts and model framework already incorporate behavioral mechanisms, future research should test this framework across a wider range of behavioral domains, which will be essential to further refine model performance.